\documentclass[runningheads]{llncs}
\usepackage{graphicx}
\usepackage{amsmath,amssymb} 
\usepackage{color}
\usepackage{makecell}
\usepackage{subcaption}
\usepackage{wrapfig}
\usepackage[linesnumbered,ruled,vlined]{algorithm2e}
\usepackage[table,xcdraw]{xcolor}
\usepackage[pagebackref,breaklinks,colorlinks]{hyperref}
\usepackage[width=122mm,left=12mm,paperwidth=146mm,height=193mm,top=12mm,paperheight=217mm]{geometry}
\begin{document}
\pagestyle{headings}
\mainmatter

\title{Practically adaptable CPABE based Health-Records sharing framework} 

\titlerunning{CPABE \& OAuth2.0}

\authorrunning{Imam et al.}

\author{Raza Imam, Faisal Anwer}
\institute{Aligarh Muslim University}

\maketitle

\begin{abstract}
With recent elevated adaptation of cloud services in almost every major public sector, the health sector emerges as a vulnerable segment, particularly in data exchange of sensitive Health records, as determining the retention, exchange, and efficient use of patient records without jeopardizing patient privacy, particularly on mobile-applications remains an area to expand. In the existing scenarios of cloud-mobile services, several vulnerabilities can be found including trapping of data within a single cloud-service-provider and loss of resource control being the significant ones. In this study, we have suggested a CPABE and OAuth2.0 based framework for efficient access-control and authorization respectively to improve the practicality of EHR sharing across a single client-application. In addition to solving issues like practicality, data entrapment, and resource control loss, the suggested framework also aims to provide two significant functionalities simultaneously, the specific operation of client application itself, and straightforward access of data to institutions, governments, and organizations seeking delicate EHRs. Our implementation of the suggested framework along with its analytical comparison signifies its potential in terms of efficient performance and minimal latency as this study would have a considerable impact on the recent literature as it intends to bridge the pragmatic deficit in CPABE-based EHR services.


\end{abstract}

\section{Introduction}
Electronic Health uses digital resources to transcend geographical obstacles and provide access to healthcare solutions. E-Health integration has been thought to be especially advantageous in enhancing the customer experience and avoiding basic clinical resources and costs. By leveraging cloud-based services and distributing healthcare data in telemedicine is like Electronic Health Records (EHRs), Electronic Medical Records (EMRs), or Personal Health Records (PHRs), practitioners and higher governing authorities may exchange the system accurate diagnoses in less time and give higher quality care to patients as a result.  Healthcare data in the electronic health record comprises personal and sensitive information that fraudsters may find appealing \cite{guo2019flexible}. As a result, one major problem for electronic health services is determining how to retain, exchange, and use patient records without jeopardizing patient privacy \cite{wang2012implementing}. Furthermore, not only must the anonymity and integrity of clinical information be safeguarded from unauthorized parties, but also against illegal access attempts from within the electronic health record frameworks (e.g., authority and CSP).

\begin{figure}[!t]
    \centering
    \begin{subfigure}[b]{0.28\textwidth}
        \includegraphics[width=\textwidth]{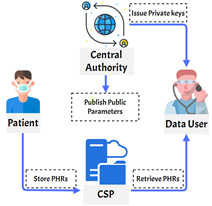}
        \caption{EHR}
        \label{fig:ehr}
    \end{subfigure}
    \begin{subfigure}[b]{0.45\textwidth}
        \includegraphics[width=\textwidth]{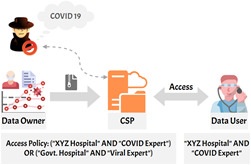}
        \caption{EHR exploitation}
        \label{fig:exploit}
    \end{subfigure}
    \begin{minipage}[b]{0.25\textwidth}
        \caption{(a) Framework for Electronic Health Record (EHR) System. (b) Exploitation of Sensitive Health Records over Cloud.}
        \label{fig:intro}
    \end{minipage}
\vspace{-0.5cm}
\end{figure}


\subsection{Background}
The underlying fundamental of Electronic Health is the cloud service, which is commonly utilized by organizations and individuals to outsource data and has been growing in popularity as cloud computing providers deliver their customer’s low-cost data space with great flexibility and stability \cite{fugkeawdesign}. Yet, when inadequate security methods are in effect, the Health Records on the cloud might be exploited by cloud service providers or spills private sensitive data \cite{pavani2019survey} as shown in Figure \ref{fig:exploit}. Encryption is amongst the most effective approaches for protecting EHRs in the cloud as it ensures security since files are encrypted before transmission to the cloud but are decrypted following their retrieval. Conventional encryption methods, such as public-key encryption, nonetheless, only require a single individual to decrypt the content, which is often too limiting and ineffectual \cite{imam2021systematic}, \cite{anas2022elliptic}, \cite{imam2022systematic}. Throughout many current applications, the data quantity is large, and the data typically necessitates a variety of access restrictions to allow several consumers to exchange the data while each user only has access to a fraction of it. Cloud applications demand more versatile fine-grained access control to enable participants with varying access control rules to facilitate Health Data exchange (See Figure \ref{fig:ehr}). The latest advancements in cryptographic concepts, such as Attribute-Based Encryption (ABE) propose alternatives to this data exchange quandary. For consumers with their respective attribute sets in ABE, particularly Ciphertext-Policy ABE (CPABE), an encrypted form of EHR is formed with the coupling of an access policy with the original file. Multiple individuals can decipher the encrypted message if their attributes satisfy the access policy defined within the encrypted text \cite{p2018attribute}.

\subsection{Motivation}
In recent times, even being the originality given by the ABE approach has been acknowledged, the novelty is found as a theory in the research world since the practicality of the ABE system in real-world operations has received little attention. This has provided us with the motivation, as in this work we recommend a framework design and implementation specifics for using an ABE scheme in a real-world application. This proposed framework containing application server not only offers security to secure users' confidential EHRs, but it also includes a sophisticated access control structure that allows numerous consumers to exchange the data. Our proposal's objective is to provide basis for examining the best set of tactics for producing a viable ABE-based solution. This study takes a use case scenario of healthcare institutions and centers where multiple institutions including City Hospitals, University Medical Colleges, Healthcare centers, Medical Nursing Homes, and Medical research laboratories share their sensitive health data with the state and central government along with several other national and multi-national authorities for specialists and researchers in data analytics, computer science, and machine learning for further accelerating the developments that potentially tackle some of the world's most difficult challenges \cite{tao2021practical}. As in the case of recent COVID-19, rapid compilation and processing of COVID-19–related EHR data, following an extensive and rigorous data collection process from infected patients, enabled the study of chronic COVID-19 fatalities and ultimately aided in the speedy release of relevant vaccines to the public for an efficient cure and recovery. Therefore, this study aims to enhance the practicality of the CPABE framework by integrating it with Google Cloud services along with OAuth 2.0 (Open Authorization) \cite{oauthgoogle}, implementing which will directly enhance the security of  authorization up to the level of Google services. 

\subsection{Contribution}
In existing scenarios of transmitting Electronic Health Records, Cloud Service Providers should safeguard the secrecy of the Data Owner's Health Record, which could result in two significant dangers for Data Owners: data-trap inside a specific CSP and denial of resource control \cite{xhafa2015designing}. This has given us the primary motive to the proposal of this work in addition to practicality, as we propose a preventive approach to avoid two aforementioned issues, we suggest interaction between Application Service Providers and unreliable cloud storage. According to our framework proposal, sharing private data over cloud environment will indeed create a simple alternative for performing two functions simultaneously, notably specific function of the application itself, along with making the Data accessible to institutions, governments, and organizations seeking delicate Patient Information on a single platform. Our method incorporates the CPABE and OAuth 2.0 standards, as well as addressing the possible issue of data entrapment within a single CSP and the loss of resource control. The implementation of the proposed framework involving CPABE over OAuth would influence the practicality among the users and would have a significance influence as it aims to fill the practical gap in EHR services based on CPABE frameworks. 

\section{Literature Review}
\subsection{Preliminaries}
The capacity to store vast amounts of data collectively is a significant competitive benefit in cloud services. Nevertheless, these data can span in wide range of format, origin, and confidentiality degree — especially when restrictions relating to consumer data or financial details are taken into account. Coarse-grained access control, or generalized access control, may function in on-premises situations where information and data can be retained independently, and access to particular data types can merely be allocated depending on storage site (e.g., Andy can access X folder, Bill can access Y folder, etc.) \cite{kumar2013review}. However, in case when data is stored in the cloud collectively, fine-grained access control is necessary because it enables data with varied access control to remain in the same storage area without causing security or regulation difficulties. Role-Based, Attribute-Based, and Policy-Based Access Controls are three kinds utilized in appropriate frameworks \cite{kumar2013review}. Therefore, Attribute Based Encryption, which among various Fine Grained Access Control models, provides the best security and reliability across Cloud storages.
\subsubsection{Attribute}
Attribute: “Attribute” is fundamentally defined as characteristic or trait of an entity, where this entity can be anything or anyone \cite{abacwikipedia}. For instance, if a person is a student, then one attribute of this person is “student”. Attributes are subdivided into 4 divisions, which are as follows:
\begin{itemize}
    \item \textit{Subject Attributes:} Attributes specifying the user’s position or role in the access, for example, university, department, job position, age, etc.
    \item \textit{Action Attributes:} Attributes specifying the user’s actions that is being executed, for example, read, modify, delete, view, etc. 
    \item \textit{Object Attributes:} Attributes the mentions the resource or entity that is being accessed, for example, health record, university data, COVID records, etc. 
    \item \textit{Contextual Attributes:} Attributes specifying the access control paradigm concerning time, location, or any other dynamic characteristics. 
\end{itemize}


\begin{figure}[t]
    \centering
    \begin{subfigure}[b]{0.35\textwidth}
        \includegraphics[width=\textwidth]{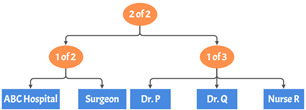}
        \vspace{0.05cm}
        \caption{$(\text{Dr. P} \lor \text{Dr. Q} \lor \text{Nurse R}) \land (\text{ABC Hospital} \lor \text{Surgeon})$, where ‘$\land$’ denotes boolean AND, and ‘$\lor$’ denotes boolean OR.}
        \label{fig:access_tree}
    \end{subfigure}
    \begin{subfigure}[b]{0.55\textwidth}
        \includegraphics[width=\textwidth]{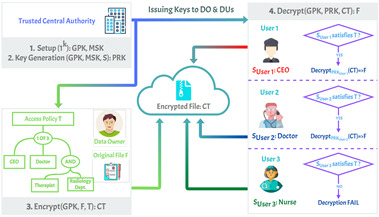}
        \caption{Ciphertext Policy ABE}
        \label{fig:CPABE}
    \end{subfigure}
    \caption{(a) An example of Access Tree representing an Access Policy. (b) Framework of standard CPABE system with entities including Trusted Authority, Data Owner, and Data User.}
\vspace{-0.5cm}
\end{figure}

\subsubsection{Access Policy}
Fine-grained access control is a way of limiting who has access to specific data. Fine-grained access control, as opposed to generic access control, also known as coarse-grained access control, employs more subtle and changeable ways for granting access \cite{kumar2015enhanced}, \cite{finegrainedaccesscontrol}. Fine-grained access control, which is most commonly utilized within cloud services, provides every data entity with its own defined access policy when vast volumes of data resources are kept together. Access Policy or Access Structure is a statement that primarily consists of various attributes to define what actions should be permitted using boolean combination (using logic and booleans ‘$\land$’ and ‘$\lor$’, where ‘$\land$’ denotes boolean AND, and ‘$\lor$’ denotes boolean OR). In Attribute Based Access Control (ABAC), access policies are essentially used for granting or revoking access to the data. Policies are also differentiated into local and global natures and can be designed in such a manner that they are capable of overriding other policies \cite{alternativesrolesclaims}. The Access Structure specifies the control over ciphertext in the case of CPABE, whereas in the KPABE case, it specifies the access scope over the decryption key of the user. Several examples could be: A hospital's data can only be viewed if the viewing user belongs to the Cardiology department, or data cannot be accessed after 6:00 p.m. ABAC allows defining as many policies that correspond to a wide range of circumstances and scenarios \cite{sowjanya2020ciphertext}.

Access trees are primarily an illustrative depiction of access policy, or in terms of ABE, when a data owner wants a data or file encrypted, he specifies the Access Policy utilizing Access Tree Structure. Considering a use-case scenario of Ciphertext Policy Attribute-Based Encryption (CPABE), if a Data Owner wants his data to be only accessed by hospital staff named Dr. P, Dr. Q, or Nurse R, who belong only to either ABC Hospital or is a Surgeon, then the Data Owner will define a specific Access Policy as: Refer to Figure~\ref{fig:access_tree}.

\subsubsection{ABE Framework}
Initially, the notion of ABE was presented by Sahai and Waters in their proposal given as fuzzy identity-based encryption \cite{goyal2006attribute}, \cite{sahai2005fuzzy}. ABE is fundamentally divided into two based on the Access Control, KPABE and CPABE. The definition of KPABE and CPABE are differentiated in relatively converse manner to a certain degree (See Fig 4). In KPABE, the encrypted data is correlated with the data owner’s attributes, while the issued Private Key is coupled with the decrypting user’s Access Policy. Data Owner or rather the encrypting party cannot authorize who could decrypt the original data, and may only specify the collection of attributes required in decryption \cite{pang2014survey}. Furthermore, the Private key which is generated by a Central Authority decrypts the original data only when the end user provides the correct attributes that comprise the Access Policy \cite{al2019survey}. On the contrary, the encrypted data, in CPABE scheme, is correlated with the Access Policy defined by the Data Owner under which the decryption can be possible, whereas the Private Key is linked with the Attributes set of decrypting users. CPABE scheme, which is incorporate in our proposal, is also illustrated in Figure \ref{fig:CPABE}. 
The algorithm of CPABE scheme is described as follows: \\


\begin{enumerate}
    
\item \textbf{Setup $(1^k)$}: \textit{Global Public Key (GPK), Master Secret Key (MSK)}\\
An authority (CA) uses this randomization method to construct a new ABE system. It involves the implicit security value $k$ as a parameter and finally provides a set of Global Public Key (GPK) as well as a Master Secret Key (MSK). As the name implies, GPK is universally known and is used in the operation of further encryption and decryption.

\item \textbf{Key Generation} $(\text{GPK}, \text{MSK}, S)$: Private Key (PRK)\\
This is used by the authority (CA) to produce a Private Key (PRK). It requires a composite of a specific user's attributes $S$, the Master Secret Key (MSK), and the Global Public Key (GPK) as arguments and subsequently yields a Private Key (PRK) for further decryption by the user with attribute $S$.

\item \textbf{Encryption} $(\text{GPK}, \text{InpFile}, T)$: Encrypted File (EncFile)\\
This stage is in charge of encrypting the input File (InpFile) with GPK and generating ciphertext, CT, in accordance with the supplied decryption policy $T$, and is processed by the Data Owner (DO). Any DO in the network can encrypt InpFile by imposing access constraints that only users who meet the access policy will be able to decode it.

\item \textbf{Decryption} $(\text{PRK}, \text{EncFile}, \text{GPK})$: Input File (InpFile)\\
The Data User (DU) performs the decryption phase, in which the DU linked with his attributes $S$ attempts to decrypt the previously encrypted file that is imposed with Access policy $T$. The original Input file InpFile can be successfully decrypted using its own issued Private key associated with Attributes and GPK following the Data User satisfies the access policy $T$.
\end{enumerate}

\subsubsection{OAuth2.0}
OAuth is a framework for delegated authorization for REST/APIs. It allows Data Users to gain restricted access (scopes) to a Data owner's data without revealing the owner's password and similar sensitive credentials. It separates authentication from permission and provides a variety of use cases that handle various device capabilities \cite{leiba2012oauth}. Server-to-server apps, browser-based apps, mobile/native apps, and consoles/TVs are all supported. In principle, OAuth is a conventional platform that enables the Data User, who is requesting for the original data, to obtain resources from another service provider or application on behalf of the Data Owner without supplying the Data User with any of the owner's credentials. OAuth is a protocol that services can utilize to deliver "secure delegated access" to client or rather Data Users. OAuth operates over HTTPS and uses access tokens instead of credentials for the authorization purposes. The workflow of the OAuth 2.0 protocol is given in the use-case as follows:
\begin{enumerate}
\item Data User registers with an application service provider, such as Facebook. 
\item Facebook assigns Data User with a “client secret” or “access token” exclusive towards that user. 
\item Data User provides “client secret” with every request for the requested Data. 
\item The request will be refused if any of the OAuth requests are invalid, lack data, or contain the incorrect token. 
\end{enumerate}
However, there are two variants of OAuth developed over time, OAuth 1.0 \cite{mosha2010rfc5849} vs OAuth 2.0 \cite{ietf2012rfc6749}. The modifications around OAuth 1.0 and OAuth 2.0 impacted the nature of OAuth in such a way that the variants really meet distinct demands depending on what one is attempting to do. Creating a reliable OAuth system is a difficult task. Although OAuth 2.0 is significantly easier to implement than OAuth 1.0 due to its cryptographic complexities, the latest version involves many vulnerability shortcomings. Most of the current services including Google and Facebook have already moved away from OAuth 1.0 in the last decade. Even so, depending on the security requirements one can choose either option in its implementation \cite{oauthdifference}. 

{\renewcommand{\arraystretch}{1.4}
\begin{table}[b]
\caption{Significant descriptive parameters of several related studies.}
\resizebox{\textwidth}{!}{%
\begin{tabular}{l|ccccc}
\hline
\rowcolor[HTML]{EFEFEF}
\textbf{Method}  & \multicolumn{1}{c}{\textbf{Focused Issue}} & \multicolumn{1}{c}{\textbf{Framework(s)}} & \multicolumn{1}{c}{\textbf{Use-Case}} & \multicolumn{1}{c}{\makecell{\textbf{Performance}\\\textbf{Analysis}}} & \multicolumn{1}{c}{\makecell{\textbf{Security}\\\textbf{Proof}}} \\ \hline
Tassanaviboon \cite{tassanaviboon2011oauth} & Encryption & OAuth and ABE & None & Yes & Yes \\
Wang et al. \cite{wang2016sieve} & Security & User-centric storage model & Health & Yes & Yes \\
Tassanaviboon \cite{tassanaviboon2011secure} & P2P Security & Self-Organizing CA group & Decision & Yes & Yes \\
Vijayan \cite{vijayan2015review} & Access Control & OAuth scheme & None & No & No \\
Jang-Jaccard \cite{jangjaccard2018practical} & Practicality & Client App \& CPABE & Generic & Yes & No \\
Ours & Practicality & Client App, CPABE, \& OAuth 2.0 & EHRs & Yes & No \\ \hline
\end{tabular}}
\end{table}}

\subsection{State Of The Art}
The early studies of implementing ABE in practical scenarios have been going on since 2011 as Tassanaviboon and Gong \cite{tassanaviboon2011oauth} introduced AAuth, a novel authorization mechanism based on the OAuth standard that uses ciphertext-policy attribute-based encryption and an ElGamal-like mask over the HTTP protocol. End-to-end encryption and ABE-based tokens are used in their method to allow permission by both authorities and owners, as well as to relocate policy enforcement from clouds to destinations. Owners can regain control of their data when it is stored in semi-trustworthy cloud storage thanks to their user-centric strategy. Furthermore, because most cryptographic operations are outsourced from owners to authorities, owners can benefit from cloud computing. Their technique retains the same degree of security as the original encryption system and safeguards users from disclosing their credentials to application providers, according to security analysis. Furthermore, Wang et al. \cite{wang2016sieve} introduce Sieve, a novel platform that exposes user data to web services selectively (and securely). Sieve uses a user-centric storage approach, in which each user uploads encrypted data to a single cloud store, with the decryption keys known only by the user by default. Sieve offers an infrastructure to support rich, legacy web applications based on this storage type. Sieve uses attribute-based encryption to allow users to design cryptographically enforced access controls that are easy to comprehend. Sieve may re-encrypt user data on storage providers in-place, revoking decryption keys from web services without exposing new keys to the storage provider, thanks to key homomorphism. 

Focusing on the semi-trusted cloud environment, Tassanaviboon \cite{tassanaviboon2011secure} is concerned with the issuance of public-key certificates in peer-to-peer (P2P) networks, which is an important basis for securing P2P functionality and applications. They developed the Self-Organizing and Self-Healing CA Group (SOHCG), which can issue certificates without the need for a centralized Trusted Third Party (TTP). A CA group is created in a Content Addressable Network (CAN) by trustworthy bootstrap nodes and then develops to maturity on its own, according to its architecture. The membership in a CA group is dynamic and has a uniform distribution across the P2P network, based on their group management policies and specified criteria; the size of a CA group is managed to a level that balances performance and acceptable security. Despite OAuth being one of the most widely used authorisation mechanisms, however in the event of heterogeneous clouds, it cannot be employed. Vijayan \cite{vijayan2015review} suggested the Fuzzy Authorization technique to solve this problem. Modified Ciphertext Policy-Attribute Based Encryption (CP-ABE) and OAuth methods are employed in the system. This approach allows an application running on one cloud party to access data stored on another cloud party. Fuzzy Authorisation is a reading authorization technique that allows only the data owner to modify the data. When a piece of data is changed, the application's right to access it is automatically revoked. This technique has become more scalable and versatile because to the inclusion of the Linear Secret Sharing Scheme and GRS code. 

In respect to untrusted cloud storages, Jang-Jaccard \cite{jangjaccard2018practical} discuss about their efficient design and deployment of a cloud storage client application that implements the concept of ABE framework. Their suggested client application has an effective access control system that allows several variations of access policies to be created, allowing numerous users to share big databases. Each user only needs to access a tiny portion of the vast data by using distinct access policies. The experiment's purpose was to find the best set of tactics for creating a viable ABE-based system. They discovered the many features and challenges related with designing a realistic ABE-based application through installation and assessment. These are several coherent research across the notion of Attribute Based Encryption and Authorization frameworks, which hold a significant scope across similar literature, and stemming from the similar notion, we provide our suggestions in the next sections. A summary of the discussed literature and their vital parameters are discussed in Table 1.

\section{Method}
\subsection{Problem Scenario}
In today's technological world of numerous mobile applications pertaining to health services, many users, or rather users operating as Data Owners, own the sensitive health records and provide those data to the client application for a range of functions, with data analysis and monitoring of daily health activities being a good example of these purposes in aspect of the growth of smart devices like smartwatches and fitness trackers using such applications. Alternatively, ArogySetu, FitBit and similar applications were also made accountable for monitoring Covid19 patients and providing such sensitive health data to government entities for a variety of objectives, including research, development, and consensus analysis.  

\begin{figure}[t]
    \centering
    \includegraphics[width=0.85\textwidth]{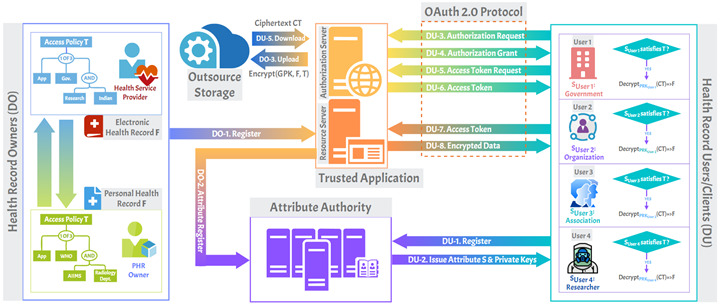}
    \caption{Architectural overview of the proposed framework incorporating the CPABE model bounded with OAuth 2.0 protocol for efficient sharing of Electronic Health Records across Multi-sharing Cloud environment. DO-(1-3) are the steps denoting the procedure in respect to the Data Owner whereas DU-(1-8) are the steps denoting the procedure in respect to the Data User.}
    \label{fig:method}
    \vspace{-0.5cm}
\end{figure}

Moreover, as cloud platforms can be considered as a sort of outsourcing method, the confidentiality of Data Owner's Health Records must be managed by Cloud Service Providers (CSP) \cite{liu2020anonymous}. As a result, this circumstance poses two critical security risks for Data Owners, the trapping of data within a single CSP, and the loss of resource control. Inter-operation involving Application Service Provider (ASP) and untrusted cloud storage is a preventative measure that can prevent customers against vendor lock-in and loss of data control. Hence, with our framework recommendation (See Figure \ref{fig:method}), this exchange of sensitive data over the cloud environment would create a simple solution for performing two operations at the same time, i.e., data analysis, monitoring, or whatever the main function of the application be, along with providing access to institutions, governments, and organizations acting as Data Users who seek sensitive Health Records over a single application. We are deploying CPABE and OAuth 2.0 frameworks throughout the client application that provides the service of health monitoring or data analysis in order to have this capability.

\subsection{CPABE meets OAuth2.0}
We have adopted the use-case scenario of an EHR sharing scheme in which an ABE scheme, specifically a CPABE framework, is implemented, which is as follows: Health records are originally registered on a trusted client-application by owners of the original data where these Data Owners could be individual users utilizing the features of the application software or respectively it could be clinicians, laboratories or diagnostic institutions holding health data in the form of PHR or interchangeably EHR. Apart from providing its specific primary services and functions, this client application would aim to provide a direct interaction between participating parties such as Data Users and Data Owners with no extra exertion on the part of the direct Data Owners as individual Data Owners would be utilizing this application only for the application's main objective whether it be monitoring, analysis, or so on, however, we have considered the main function of the application as Data analysis for convention purposes.  

The primary significance of fine-grained access control of the EHR implementation comes with the initial data itself as the data, that will be uploaded on the cloud by the application software, will have inbuilt access policy defined by the Data Owner only as the Data Owner is the original owner of the data and he should describe the access policy if who will be able to access his data. Eventually, after the Data Owner uploads his Health Record to the application, the application will upload the data by encrypting using symmetric cryptosystem, AES, which is then subsequently uploaded to the cloud storage by the application. As the encrypted data, which also holds the access policy within it, can be removed or updated by the original Data Owner meanwhile the application is conducting its main function of analysis. Furthermore, since the CPABE framework requires a Central Attribute Authority to provide the relevant attributes as well as private keys, this phase will be covered by the Application software on behalf of the Data Owner. It can also be seen that on the Data Owner's side, minimal work is being done in our framework so that the Owner of the EHR remains focused on his only intention of utilizing the application's functionality of analysis while simultaneously being aware of his data risk. Figure \ref{fig:method} depicts the architectural overview of the proposed framework incorporating the CPABE model bound with OAuth 2.0 for efficient sharing of Health Records across Cloud Environment.

Many organizations who are seeking the EHRs of the Data Owners would be interacting with the application and its components itself only instead of interacting with the Data Owner, as initially the organization requests the middleman, i.e., the application software, for the respective authorization as a Data User following the registration with the Attribute Authority with the respective attributes and relevant private keys (See Table 2). The initial authorization, after the registration with Attribute Authority, is done using OAuth Authorization, and this is where the role of OAuth 2.0 protocol comes in play. In terms of traditional client-server authentication model, there are several shortcomings and limitations that could be found like third-party services must save the data owner's privileges for future usage, or it could occur that third-party could get excessively broad access to the Data owner's protected data, disabling the owner to control period or access to a limited selection of assets. Also, any compromise with any third-party organization could also make the protected data and credentials vulnerable. The essence of OAuth would be significant here as it aims to address these issues by introducing an authorization layer and separating the role of the Data Owner from that of the third-party or Data User \cite{tassanaviboon2011oauth}, \cite{ribeiro2018implementation}. In OAuth, the Data User seeks access to the EHR, or generally protected data, owned by Data Owner and hosted by the Resource Server, and is provided with credentials that vary from Data Owner’s credentials. Rather than using the Data Owner’s credentials, the Data User acquires the Access Token, which is a character sequence specifying the scope, duration, and other access parameters. Next, an authorization server of the Application issues the respective Access Tokens to the Data Users but without the Data Owner’s approval in contrast to standard OAuth protocol where Data Owner’s approval is essential \cite{ietf2012rfc6749}. The Data Owner’s approval would not be required in our case of record sharing as the CPABE framework employs flexible Access Policy that would define if who were specifically approved of the Data Owner. The Data User then employs the Access token to the resource server and finally accessing the required electronic health records after being validated by the resource server. Access token remains valid until it expires, or specifically, until the constraints specified in the Access token expires. For having the refreshed Access token, the Data User will have to go through the process given in the OAuth 2.0 specification \cite{ietf2012rfc6749}. Refresh tokens, unlike access tokens, are solely designed for usage with authorization servers and are never provided to resource servers. In this manner, organizations or Data Users, following the issuance of private keys and attribute set by Attribute Authority along with authorization by the Client Application, will be able to access the required Health Record. Data User will consequentially be able to decrypt the initially encrypted EHR with the condition if their attribute set satisfies the defined Access Policy bounded with the EHR.

{\renewcommand{\arraystretch}{1.7}
\begin{table}[t]
\caption{Description of various components and respective responsible entities involved in the interaction of the Data Owner, Data User, and the Client Application.}
\fontsize{20}{20}\selectfont
\resizebox{\textwidth}{!}{%
\begin{tabular}{lll}
\rowcolor[HTML]{EFEFEF} 
\hline
\textbf{Component} & \multicolumn{1}{c}{\cellcolor[HTML]{EFEFEF}\textbf{Description}} & \multicolumn{1}{c}{\cellcolor[HTML]{EFEFEF}\textbf{Responsible entity}} \\ \hline
Health Record & Electronic copies of a patient's medical and treatment data & Data Owner \\
Role: Data Owner & Patients themselves, physicians, or medical team holding EHRs & Data Owner \\
Role: Data User & Institutions, governments, organizations, researchers, etc. & Data User \\
Role: Authorization Server & Server responsible for issuing Access Tokens following Authorization & Application \\
Role: Resource Server & Receive, maintain, \& react to EHR requests using Access Token & Application \\
Attribute & Role, characteristic or trait of an entity or user & Owner/User \\
Access Policy & Structured condition defining what actions should be permitted & Data Owner \\
Authorization Grant & Credential representing Resource Owner's permission to access EHRs & Application/User \\
Access Token & Credential indicating issued authorization, \& for accessing EHRs & Data User \\ \hline
\end{tabular}}
\vspace{-0.5cm}
\end{table}}

\begin{figure}[t]
\centering
    \begin{minipage}{0.6\textwidth}
        \begin{subfigure}{\textwidth}
            \includegraphics[width=\textwidth]{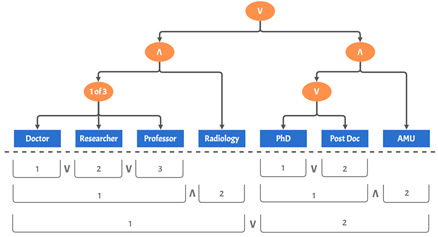}
        \end{subfigure}%
    \end{minipage}
    \begin{minipage}{0.29\textwidth}
        \caption{Access Policy Tree representing the Attribute Policy defined by to Data Owner in relation to the proposed use-case scenario of Electronic Health Record.}
        \label{fig:exp_accesspolicy}
    \end{minipage}
    \vspace{-0.5cm}
\end{figure}

\subsection{Implemented Solution}
\label{sec:solution}
In the procedure highlighted in the following, we have utilized the access policy, attribute set, and access token utilized during the implementation of the proposed framework. Each step also highlights the correspondence with respect to the Figure \ref{fig:method} framework with the notation of DO or DU or AA, where they denote the procedure in respect to the Data Owner, Data User and Attribute Authority respectively. The following describes the use-case procedure of EHR access over the proposed CPABE and OAuth framework in the procedural manner: 

\begin{enumerate}
    \item \textbf{AA:} Initially, the operation is initialized by the client application itself as it initializes the Setup phase of the CPABE framework by inputting the security parameter $\lambda$, although in our case of EHR implementation, a random number is generated utilizing the implemented cryptographic library. \\
    
    \item \textbf{DO:} Meanwhile, the Data Owner, wishing to utilize the client application for its primary features (let’s conventionally consider this feature to be Data Analysis), registers with his attributes and uploads his Health Record (denoted F) to the Resource Server of the Application, along with the defined Access Policy while being aware that his or her data will be shared with the Data Users satisfying the specified policy. An example of Access Policy suitable to this use-case is shown in Figure \ref{fig:exp_accesspolicy}. 
    The Access Policy (denoted as \texttt{T}) defined by Data Owner where Data User has been taken University members \cite{porwal2017implementation}, which we have also used in our simulation stage is as follows:
    \begin{sloppypar}
    \texttt{T = Position: Doctor Position: Researcher Position: Professor 1of3 Department: Radiology 2of2 Position: PhD Position: Postdoc 1of3 University: AMU 2of2}\\
    \end{sloppypar}
    In general, boolean form of the above Access Policy can be as 
    \begin{sloppypar}
    \texttt{T = [[[[position = Doctor OR position = Researcher OR position = Professor] AND Department = Radiology] OR position = PhD OR position = Postdoc] AND University = AMU]\\ }
    \end{sloppypar}

    \item \textbf{DO:} Since CPABE framework requires Attribute Authority for managing and distributing private keys and attributes, hence the Client application registers the Attributes of the Data Owner with the Attribute Authority. \\

    \item \textbf{AA:} As the encrypted data gets uploaded by the Client Application on the Cloud using AES-256 Asymmetric algorithm, the organizations and data seekers depicted as Data Users, initially registers with the Attribute Authority in order to have the relevant Attribute set(s) for itself, followed by the generation of private key as per the attributes. For example, several simple attributes set (denoted as S) of 3 respective Data Users be S0, S1, and S2, are as: 
    \begin{sloppypar}
        \texttt{S0 = Position: Doctor, Department: Radiology, University: AMU 
            S1 = Position: Student, University: AMU 
            S2 = Position: PhD, College: JNMC", University: AMU\\
            }
        \end{sloppypar}

    \item \textbf{DU:} Now, Authorization over OAuth 2.0 protocol will be performed as Data User initially requests for authorization from the authorization server. Then, Data User is provided with an authorization grant, that is an identity credential signifying the authority of the Data Owner for accessing the particular Health Record. \\

    \item \textbf{DU:} The role of Access Token, the most significant component of OAuth protocol, plays its role here as the Data User requests an access token as soon as the Authorization server authenticates the presented Authorization grant. This access token contains the relevant scope, duration, and other access parameters for the Data Access. \\

    \item \textbf{DU:} If the authorization grant presented by Data User turns out legitimate, the authorization server authenticates Data User by issuing an access token. \\

    \item \textbf{DU:} Following the authentication utilizing the access token, Data User queries the resource server for the encrypted EHR File while the access token gets validated by the resource server, and if turns legitimate, the request is served. The Data User will consequentially be able to decrypt the initially encrypted EHR but only if its attribute satisfies the Access Policy. 

\end{enumerate}

\section{Simulation Analysis}
The suggested framework that utilizes the CPABE protocol and OAuth 2.0 framework utilizes the JAVA realization for CPBAE based on the library Java Pairing Based Cryptography Library (JPBC) \cite{cpabegithub}, while for the Open Authorization 2.0, we utilized the API provided by Google along with the Google Drive Storage as a medium of Cloud Storage. CPABE and OAuth are implemented on the JAVA class \cite{oauthgoogle}. In terms of hardware specifications that are used in implementation and analysis are quad CPU with clock rate of 2.1-3.7 GHz AMD Ryzen 5 3500U alongside 8 GB of RAM. However, we extracted the data from Jaccard's implementation from their work without implementing their framework on our system. Nonetheless, we found our system configuration to be identical to Jaccard's, with the exception of RAM, as our configuration utilizes 8 GB while theirs uses 16 GB of RAM. We also discovered a distinguishing pattern. In the performance comparison phases, where our model showed significantly better results even with 8 GB of RAM, it is obviously apparent that it will perform better with 16 GB of RAM as well. As a result, we can conclude that there is no discrepancy in the configuration, implementation, and comparison of the two methods. This section discusses the analysis of various simulations of our framework in terms of encryption, decryption, authorization, access policies, and attributes in comparison to a similar approach of Jaccard \cite{jangjaccard2018practical}. 

{\renewcommand{\arraystretch}{1.3}
\begin{table}[t]
\caption{Various Attributes incorporated with the respective Data Users in our simulation of implementation design.}
\resizebox{\textwidth}{!}{%
\begin{tabular}{ccc}
\hline
\rowcolor[HTML]{EFEFEF} 
\textbf{\# Attributes} & \multicolumn{1}{c}{\cellcolor[HTML]{EFEFEF}\textbf{Associated Attributes of Data Users}} & \multicolumn{1}{c}{\cellcolor[HTML]{EFEFEF}\textbf{Satisfies Access Policy}} \\ \hline
1 & \texttt{Position: Doctor} & No \\
2 & \texttt{Position: PhD, University: AMU} & Yes \\
\vspace{0.2cm}
3 & \texttt{Position: Doctor, Department: Radiology,  University: AMU} & Yes \\ 
\vspace{0.2cm}
4 & \makecell{\texttt{Position: PhD, College: JNMC, University: AMU,}\\ \texttt{Department: Radiology, City: Aligarh}} & Yes \\
\vspace{0.2cm}
5 & \makecell{\texttt{Position: Researcher, University: AMU,}\\ \texttt{Department: Radiology, City: Aligarh, College: JNMC}} & Yes \\
\vspace{0.2cm}
6 & \makecell{\texttt{Position: Postdoc, College: JNMC, University: AMU,}\\ \texttt{Department: Radiology, City: Aligarh, Position: Researcher}} & Yes \\
\vspace{0.2cm}
7 & \makecell{\texttt{Position: PhD, College: JNMC, University: AMU,}\\ \texttt{Department: Radiology, City: Aligarh, Position: Researcher,}\\ \texttt{Status: Temporary}} & Yes \\
\vspace{0.2cm}
8 & \makecell{\texttt{Position: Doctor, College: JNMC, University: AMU,}\\ \texttt{Department: Radiology, City: Aligarh, Position: Researcher,}\\ \texttt{Status: Permanent, Year: 2022}} & Yes \\ \hline
\end{tabular}}
\vspace{-0.5cm}
\end{table}}

\subsection{Performance Analysis}
We have performed the Key generation, Encryption, and Decryption phases of our proposed framework, and eventually compared them with the similar work of Jang-Jaccard \cite{jangjaccard2018practical}. Jaccard also focused on the practical approach using ABE framework, and for convention purposes, we have denoted this compared work using ‘Jaccard’ throughout our simulation. The execution time (in millisecond) of every phase of the CPABE framework depends vastly on the number of attributes any Data Owner or Data User is associated with, and hence, this characteristic can be utilized in measuring and comparing the performance of the proposed framework. In our simulation, we have considered the scenario, where several users belonging to a Medical College, JNMC, of Aligarh Muslim University, are acting as Data Users, and are requesting the previously stored Health Records to the Client Application, where each Data User have their specific set of attributes. Table 3 also shows the DU-specific attributes we have designed over the defined Access Policy (shown in Figure \ref{fig:exp_accesspolicy} and Section \ref{sec:solution}) for our EHR use-case simulation. 

\begin{figure}[t]
    \centering
    \begin{subfigure}[b]{0.45\textwidth}
        \includegraphics[width=\textwidth]{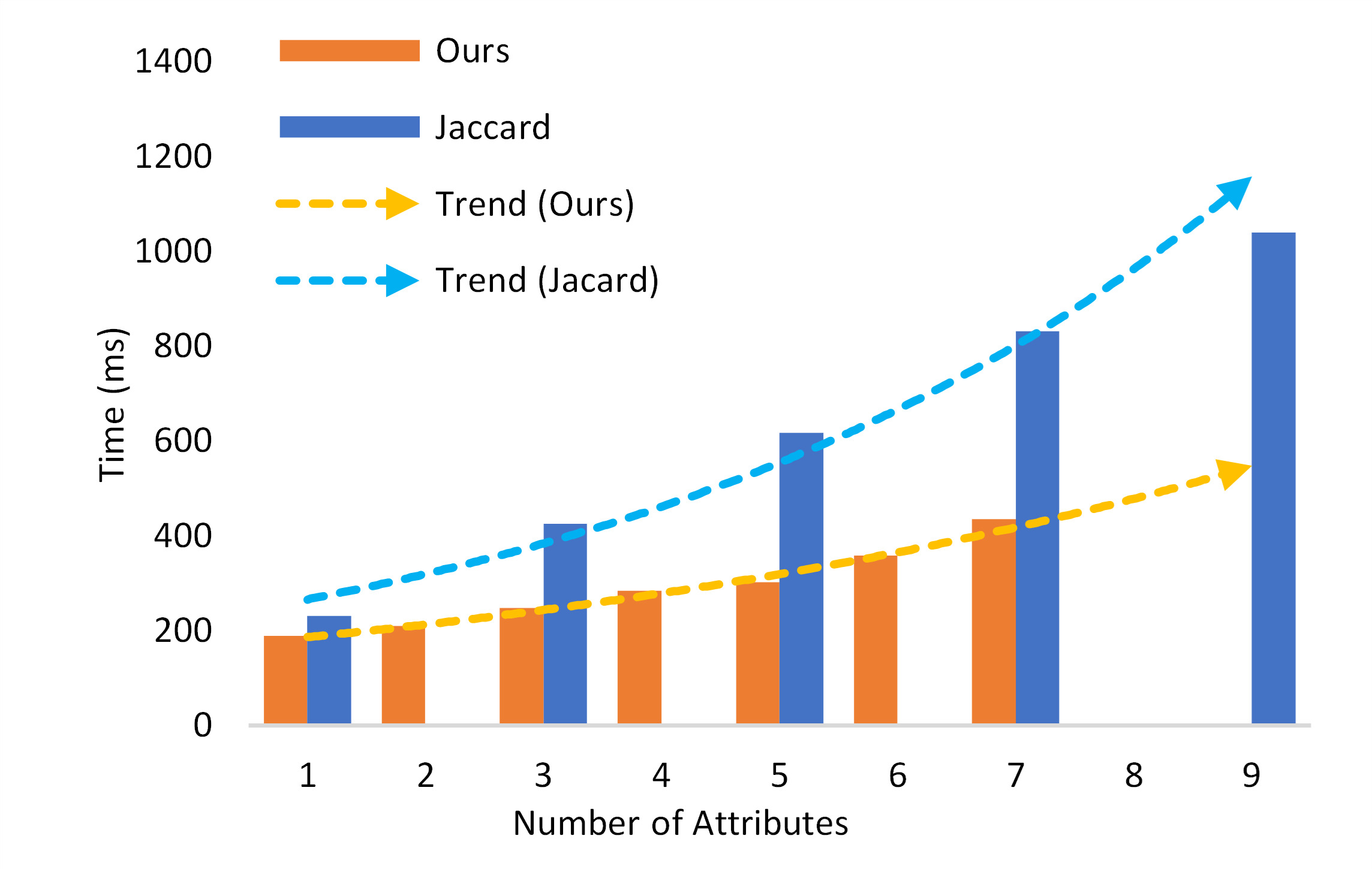}
        \caption{Key Generation}
        \label{fig:keygen}
    \end{subfigure}
    \begin{subfigure}[b]{0.45\textwidth}
        \includegraphics[width=\textwidth]{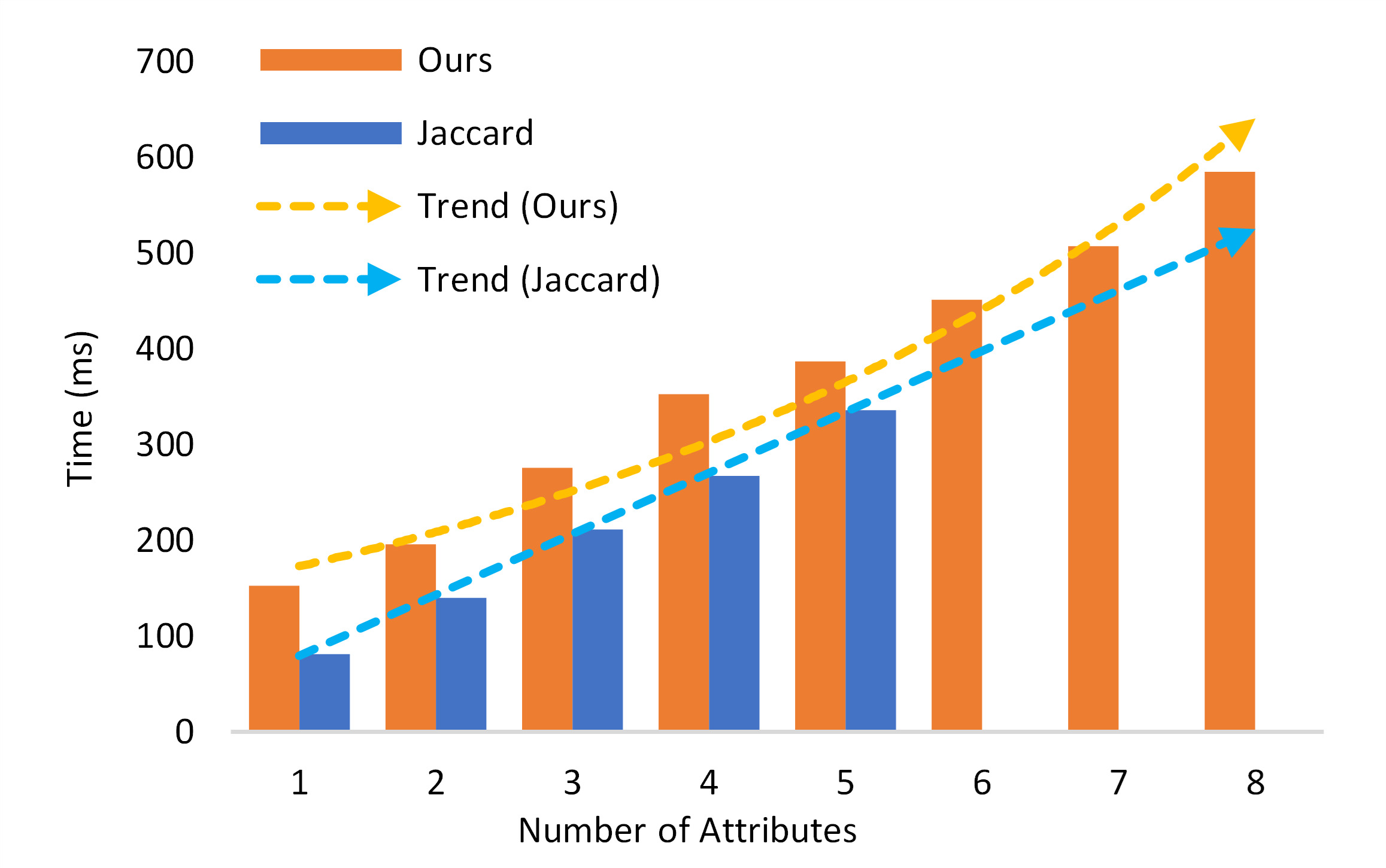}
        \caption{Encryption}
        \label{fig:enc}
    \end{subfigure}
    \caption{Final Caption. (a) Comparison of our scheme and Jaccard Scheme in terms of Encryption cost under 10 attributes. (b) Comparison of our scheme and Jaccard Scheme in terms of Decryption cost under 10 attributes.}
\vspace{-0.5cm}
\end{figure}

\subsection{Key Generation}
\begin{figure}[t]
    \centering
    \begin{subfigure}[b]{0.45\textwidth}
        \includegraphics[width=\textwidth]{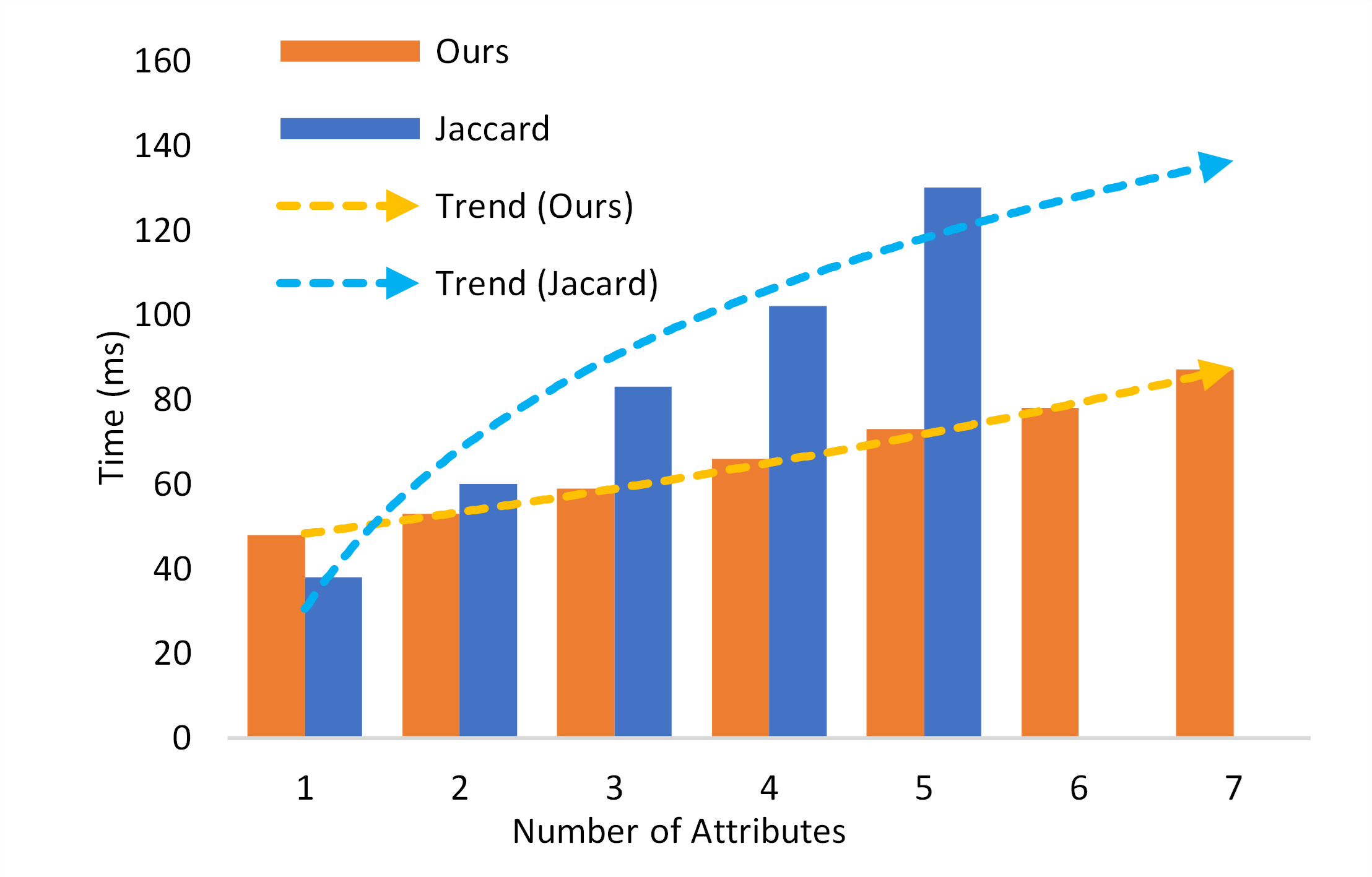}
        \caption{Decryption}
        \label{fig:dec}
    \end{subfigure}
    \begin{subfigure}[b]{0.45\textwidth}
        \includegraphics[width=\textwidth]{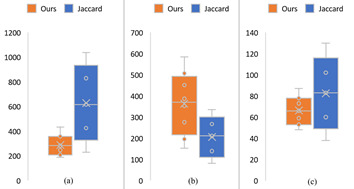}
        \caption{Overall Efficiency}
        \label{fig:boxplot}
    \end{subfigure}
    \caption{(a) Comparison of our scheme and Jaccard Scheme in terms of Decryption cost under 10 attributes. (b) Comparison of our scheme and Jaccard Scheme in terms of (a) Key Generation, (b) Encryption, and (c) Decryption cost under 10 attributes in respect to the median and outlying values on the box plots. The vertical axis on the left of box plot denotes the execution time in ms (millisecond).}
\vspace{-0.5cm}
\end{figure}

In our analysis, we found that the overall efficiency of our framework showed better results in terms of key generation, and decryption phases, however, encryption phase showed a slightly lower efficiency in comparison. In the generation of keys in CPABE framework, private keys are associated with the attributes of the user for whom the keys are generated, hence the total number attributes play a major role in determining the key generation. As shown in Figure \ref{fig:keygen}, which depicts the Key generation cost in terms of required time of various Data User and Data Owners with attributes having 10 or less attributes. Although it is possible to for a user to have more than 10 attributes, but we have considered under 10 attributes for efficient and convenient simulation. Over the total of 10 attributes, our framework showed a slightly incremented pattern, or rather a linear trend, in comparison to Jaccard’s, which showed somewhat an exponential increment, as their framework showed higher generation time at every incremented number of attributes. Hence it can be said that our framework showed lower latency in terms of key generation as it can also be depicted in the Figure \ref{fig:boxplot}(a) where box plot depicts the outlying data values, along with the median value of our framework shows greater efficiency in latency. 

\subsection{Encryption Phase}
Considering the encryption phase, where the Application software acting as Client Application encrypts the original record of Data User before uploading it to the cloud storage, our framework showed a slightly low performance over maximum of 10  attributes (Figure \ref{fig:boxplot}(b)), but despite that the incremented trend of our framework showed a linear trend as depicted in Figure \ref{fig:enc}. In contrast, data generated from Jaccard’s framework depicted a slight exponential trend depicting that, in the cases where there are large number of attributes associated with a Data User, there are chances for the framework to have a higher latency in terms of encryption as compared to our framework showing linearly trend. Thus, it can also be concluded that our recommendation would perform with higher efficiency over the large number of attributes. The key generation and encryption phases take place before the Data User requests for the Health Record, hence Open Authorization, i.e., OAuth 2.0 protocol is executed before the decryption phase as mentioned in Section 3 as well.  

\subsection{Decryption Phase}
Following the authorization, Decryption phase takes place if the Data User gets succeeded in satisfying the Access Policy, defined by the Data Owner, using his registered Attributes. Following the OAuth protocol, whose workflow in our framework has been described in Section 3, the decryption phase takes place which is executed by the respective Data Users. Our simulation showed similar results for decryption phase like in the case of key generation and encryption phases, as the executed datapoints of overall executed time in respect to number of attributes showed better performance in our framework than Jaccard’s as shown in Figure \ref{fig:dec}. In contrast to our framework that followed the linear incremental trend over the maximum of 10 attributes of the Data Users, Jaccard’s scheme followed the shift towards logarithmic trend, which would have negative effect as soon as the number of attributes increases. The box plot illustrated in Figure \ref{fig:boxplot}(c) depict the median values and quartile ranges of both the compared frameworks, and signify that outlying values as well as the median values of our framework lies close within over each of the 10 simulated attributes, hence implying a lower latency, and therefore better performance in decryption compared to Jaccard’s framework.

\section{Discussion}
The performance of our suggested framework is compared and discussed with the Jaccard’s \cite{jangjaccard2018practical} framework in terms of key generation, encryption, and decryption phases. In each of the aforementioned phases, our framework performed with higher efficiency over the large number of attributes. Nevertheless, in encryption phase, despite having a slightly lower performance of our scheme, the potential performance over higher number of total attributes result in better efficiency in comparison. In key generation and decryption phase, our scheme showed higher efficiency and lower latency in comparison to Jaccard’s, which signifies its potential in terms of efficiency.
The broad implementation of cloud technology in the preceding decade has developed multiple possibilities for new solutions to flourish in every subject as a vast number of studies has been produced since then. The increased utilization of cloud services in electronic health services have provided with a variety of helpful applications including EHR sharing, but in contrast, it has also elevated numerous vulnerable scenarios in terms of reliability and privacy, hence, cloud solutions must ensure dependability and intact reliability. In recent times, multitudes of mobile-cloud services leverage on the EHRs of several Data Owners to deliver a variety of public services and capabilities such as data analysis and monitoring. In existing scenarios of transmitting Electronic Health Records, the privacy of the Data Owner's Health Record should be handled by Cloud Service Providers, which could result in two key vulnerabilities for Data Owners: data-trap inside a particular CSP and loss of resource control. 

\section{Conclusion and Future Work}
In this study, we propose an inter-operation between Application Service Providers and unreliable cloud storage as a preemptive approach to avoid clients from vendor lock-in and loss of data control. This sharing of private data over the cloud architecture, according to our framework suggestion, would create a simple solution for conducting two functions at the same time, namely the specific activity of the application, as well as making the Data accessible to institutions, governments, and organizations who seek delicate Patient Information over a single platform. CPABE and OAuth 2.0 standards have been included within our scheme, which also addresses the potential issue of data entrapment within a single CSP and the loss of resource control. Furthermore, our implementation analysis compares our approach to a similar study, and we discovered that our framework performed more efficiently across a large number of attributes. Nonetheless, in the encryption process, amidst our technique having marginally lower performance for under 10 attributes, the potential performance over a larger number of total attributes results in improved efficiency in relation. In the key generation and decryption phases, our method outperformed Jaccard's in terms of efficiency and latency, indicating its potential viability. 

\textbf{Future Work.} We intend to use the framework incorporating the Blockchain and Smart Contract frameworks in future work because of their capabilities such as immutability, decentralization, consensus, and rapid transactions. Using Blockchain in our proposed architecture, we might address challenges in EHR sharing such as single points of failure, while also improving practical adaptability in terms of anonymity and decentralization. Among comparable approaches in the ABE and Healthcare sectors, this strategy would offer the most substantial gains in terms of efficiency, as well as practical and scalable applicability in other real-world applications.

{\renewcommand{\arraystretch}{1.3}
\begin{table}[]
\caption{Description of in-text Notations and Abbreviations.}
\resizebox{\textwidth}{!}{%
\begin{tabular}{cr|cr}
\hline
\rowcolor[HTML]{EFEFEF} 
\textbf{Notations} & \multicolumn{1}{r|}{\cellcolor[HTML]{EFEFEF}\textbf{Description}} & \cellcolor[HTML]{EFEFEF}\textbf{Notations} & \multicolumn{1}{r}{\cellcolor[HTML]{EFEFEF}\textbf{Description}} \\ \hline
ABE & Attribute Based Encryption & EMR & Electronic Medical Records \\
API & Application Programming Interface & GPK & Global Public Key \\
ASP & Application Service Provider & KPABE & Key-Policy Attribute-Based Encryption \\
CAN & Content Addressable Network & MSK & Master Secret Key \\
CPABE & Ciphertext Policy Attribute-Based Encryption & OAuth & Open Authorization \\
CSP & Cloud Service Provider & P2P & Peer to Peer \\
DO & Data Owner & PHR & Personal Health Records \\
DU & Data User & PRK & Private Key \\
AA & Attribute Authority & SOHCG & Self-Organizing and Self-Healing CA Group \\
EHR & Electronic Health Records & TTP & Trusted Third Party \\ \hline
\end{tabular}}
\end{table}}

\bibliographystyle{splncs}
\bibliography{egbib}
\end{document}